\documentstyle[epsfig,aps,preprint]{revtex}

\newlength{\dinwidth}
\newlength{\dinmargin}
\def\QL{(Q\cdot L)}
\def\PL{(P\cdot L)}
\def\QN{(Q\cdot N)}
\def\PN{(P\cdot N)}
\def\NPLQ{(\widetilde{NPLQ})}
\def\LQN{(\widetilde{LQN})}
\def\QPL{(\widetilde{QPL})}
\def\PLQ{(\widetilde{PLQ})}
\def\LPN{(\widetilde{LPN})}

\def\ABC{(\widetilde{ABC})}
\def\ABCD{(\widetilde{ABCD})}
\def\bea{\begin{eqnarray}}
\def\eea{\end{eqnarray}}
\def\nn{\nonumber}
\def\g{\gamma}
\begin{document}

\preprint{      HUPD-9914}
     \preprint{     YUMS 00-001}
     \preprint{   hep-ph/0001151}

\title{Azimuthal Angle Distribution   \\
in $B \to K^* (\to K \pi) \ell^+ \ell^- $  
at low invariant $m_{\ell^+ \ell^-}$ Region  }

\author{
C.S. Kim${~^a}$\footnote{e-mail: kim@cskim.yonsei.ac.kr,
~~ http://phya.yonsei.ac.kr/\~{}cskim/}~, ~~
Yeong Gyun Kim${~^a}$, ~~ Cai-Dian L\"u${~^b}$\footnote{e-mail: 
lucd@theo.phys.sci.hiroshima-u.ac.jp,  ~http://www.geocities.com/Hotsprings/Falls/8341/}, 
~~ Takuya Morozumi${~^b}$\footnote{e-mail: 
morozumi@theo.phys.sci.hiroshima-u.ac.jp} }

\address{
$a:$ Dept. of Physics, Yonsei University, Seoul 120-749, Korea \\
$b:$ Dept. of Physics, Hiroshima University,  Higashi-Hiroshima, Japan }


\maketitle

\begin{abstract}

We present the angular distribution of the rare $B$ decay,
$B \to K^* (\to K \pi) \ell^+ \ell^-$.
By studying the azimuthal angle distribution 
in the low invariant mass region of dileptons,
we can probe new physics effects efficiently. In particular, 
this distribution is found to be quite sensitive to the ratio of the
contributions from two independent magnetic moment operators,
which also contribute to $B \to K^* \gamma$. 
Therefore, our method can be very useful 
when new physics is introduced without changing the total decay rate 
of the $b \to s \gamma$.
The angular distributions are compared with
the predictions of the standard model, and
are shown for the cases when the afore-mentioned ratio is different 
from the standard model prediction. 

\end{abstract}



\section{Introduction}

Rare $B$ decays are suitable for testing the standard model (SM) 
and the models beyond the SM. 
Exclusive decay $B \to K^* \gamma$ and 
the corresponding inclusive decay $B \to X_s \gamma$ 
place strong constraints on the parameters of 
models beyond the SM, for example,
the left-right symmetric model (LRSM), SUSY, the multi-Higgs
doublet model,  {\it etc.} \cite{bsg,B2}.
However, if the decay rate is {\it  not changed drastically} from the
prediction of the SM, it would be very difficult
to probe new physics effects from the $B \to K^* \gamma$ decay.
In this regard, new methods have been proposed, 
which consist of observables sensitive to chiral structure, 
such as
mixing-induced CP asymmetry in $B_{d,s} \to M^0 \gamma$ decay \cite{gronau}
and $\Lambda$ polarization in the $\Lambda _b \to \Lambda \gamma$
decay \cite{mannel}.   And
these methods have been also applied to search for the new physics,
as shown in \cite{hou,kim}:
The $B \to K^* \gamma$ decay occurs through the effective interaction of
two magnetic moment operators,
\bea
 m_b ( C_{7L} {\bar s}_L \sigma_{\mu \nu} b_R F^{\mu \nu}
+C_{7R} {\bar s}_R \sigma_{\mu \nu} b_L F^{\mu \nu})~.
\eea
In the SM, the first term is dominant and the second term
is suppressed by ${\cal O} (m_s/m_b)$. In the LRSM, the
contribution of both operators can be equally
important \cite{B2}. The new contributions for $C_{7R}$ and $C_{7L}$ 
in this model are enhanced as $m_t/m_b$.
Because the probability for $B$ meson decaying to 
left-handed (or right-handed) circular polarized $K^*$
is proportional to $|C_{7L}|^2$ (or $|C_{7R}|^2$), 
the polarization measurement of $K^*$ and $\gamma$ is useful for
extracting the ratio of $\frac{|C_{7L}|}{|C_{7R}|}$.
However, since the polarizations of high energy real photon ($\gamma$) 
cannot be measured easily, we have to develop more elaborated method 
for extracting the afore-mentioned ratio.
Therefore, we propose another new method, which is very efficient when
we cannot find the new physics effects from the total decay rate of 
$B \to K^* \gamma$. 

Let us imagine the decay configuration when 
$K^*$ from the decay $B \to K^* \gamma^*$ is emitted to the direction of 
$+z$ and $\gamma^*$ is emitted to the opposite direction in the
rest frame of $B$ meson. Here $\gamma^*$
is  off-shell photon and it further decays into $\ell^+  \ell^-$, 
and $K^*$  subsequently decays into $K  \pi$. If we ignore small
mixture of the longitudinal component, 
the angular momentum of $K^*$ is either $J_z=+1$ or $J_z=-1$, 
and the corresponding production amplitude 
is proportional to $C_{7R}$ or $C_{7L}$, respectively.
Suppose the final $K$ meson is emitted to
the direction of $(\theta_K,~ \phi)$ in the rest frame of $K^*$, where
$\theta_K$ is a polar angle and $\phi$ is an azimuthal  angle
between the decay plane of ($K  \pi$) and the decay plane of ($\ell^+  \ell^-$).
The decay amplitude for the whole process is proportional to 
$$ A~ C_{7L} \exp( -i  \phi) + B~ C_{7R} \exp(+i \phi)  + C~.$$ 
Here, $A$, $B$ and $C$ are the real functions of the other angles and $C$ 
corresponds to the amplitude for the $B$ meson decaying into the 
longitudinally polarized  $K^*$,
which is possible only for the off-shell photon. By squaring 
the amplitude, we can show that in the azimuthal distribution  the
coefficient of $\cos(2 \phi)$ (and that of $ \sin( 2\phi)$) is 
$Re (C_{7R} C_{7L}^*)$ (and $Im (C_{7R} C_{7L}^*)$).
Therefore, from the angular dependence we may extract 
the ratio $\frac{C_{7L}}{C_{7R}}$. Note that for on-shell photon 
the dependence on the azimuthal angle $\phi$
does not appear for the $B \to K^* (\to K \pi)+ \gamma$ decay 
because of rotational 
symmetry of the decay configuration with respect to $z$-axis.
This naturally leads us to investigate the angular distribution of the 
$B \to K^*(\to K \pi)+ \gamma^*(\to \ell^+ \ell^-)$. 
However, once we consider off-shell photon,
some complications arise and the argument discussed above has to be modified:
The other diagrams like box and $Z$ penguin diagrams now contribute to 
the same final state through the process, 
$B \to K^*(\to K \pi)+ \ell^+ +\ell^-$.
They do not contribute to $B \to K^* \gamma$.
Though in the low invariant mass region of dileptons
the decay through the magnetic moment interactions may be dominant, we 
still have to take into account  the
effect of the box and $Z$ penguin diagrams, which
have the form of the local four-fermi interactions
in the effective Hamiltonian.

The paper is organized as follows: In section 2, we derive the
angular distribution formulae in terms
of the helicity amplitudes. In section 3, our numerical analyses for 
azimuthal angle are shown. Concluding remarks are also in section 3.

\section{Angular Distribution of $B \to K^*(\to K \pi)+ \ell^+ +\ell^-$}

The short distance contribution to 
decay $B \to K^* (\to K \pi) \ell^+\ell^-$ is governed by the 
quark level decay $b\to s \ell^+\ell^-$ as
\bea
{\cal M} (b\to s \ell^+\ell^-) = \frac{G_F \alpha }{\sqrt{2}\pi }V_{ts}^*V_{tb}
              &\left[\right.& ( C_9^{eff} - C_{10})~\bar{s}_L \gamma_\mu b_L 
                           ~\bar{\ell}_L \gamma^\mu \ell_L  \nn \\
              &+&  ( C_9^{eff} + C_{10})~\bar{s}_L \gamma_\mu b_L  
                           ~\bar{\ell}_R \gamma^\mu \ell_R  \nn \\
          &-& 2 C_{7L}   ~\bar{s_L} i \sigma_{\mu \nu } \frac{L^\nu}{L^2}                          m_b   b_R ~\bar{\ell} \gamma^\mu \ell \nn \\
          &-& \left.2 C_{7R}  ~\bar{s_R} i \sigma_{\mu \nu } \frac{L^\nu}{L^2}                          m_b   b_L ~\bar{\ell} \gamma^\mu \ell ~\right] ,
\eea
where we assume that new physics effect does not
change the  
Wilson coefficients $C_9$ and $C_{10}$ and only can change 
the coefficients of non-local four-fermi interactions which are 
denoted by $C_{7}$. The latter also
contributes to $b \to s \gamma$. 
In the SM, $C_{7L}=C_{7eff}, C_{7R}=\frac{ms}{mb}C_{7eff}$,
where $C_{7eff}$ is given in Ref. \cite{buras}. 
Although there are overwhelming resonance contributions from $J/\psi$ and
$\psi'$, {\it etc.}, the short distance contribution still dominates the low
invariant mass region of the lepton pair \cite{long}.
The effective Hamiltonian for the corresponding $b\to s \ell^+\ell^-$
is \cite{buras}
\bea
{\cal H}_{eff}(b\to s \ell^+\ell^-)&=&{\cal H}_{eff}(b \rightarrow s \gamma)
-\frac{G_F}{\sqrt{2}} V_{tb}V_{ts}^* {\alpha \over \pi}
\sum _{i=9}^{10} C_i O_i, \nn \\
{\rm where}~~~~~{\cal H}_{eff}(b \rightarrow s \gamma)&=&
-\frac{G_F}{\sqrt{2}} V_{tb}V_{ts}^* 
[C_{7L}O_{7L}+C_{7R}O_{7R}].
\eea
The operators $O_i$ relevant for us are
\bea
  O_9&=&({\bar s} b)_{L} ({\bar \ell} \ell)_{V},\\
  O_{10}&=&({\bar s} b)_{L} ({\bar \ell} \ell)_{A},\\
O_{7L} &=& 
\frac{em_b}{4 \pi^2} (\bar s_L \sigma_{\mu \nu} b_R) F^{\mu\nu},\\
O_{7R} &=& 
\frac{em_b}{4 \pi^2} (\bar s_R \sigma_{\mu \nu} b_L) F^{\mu\nu},
\eea
where in addition to the SM operators $O_9$, $O_{10}$ and  $O_{7L}$,
we include also a new operator $O_{7R}$. The new physics effects can contribute
to any of the operators.
For example, the LRSM \cite{LRSM} 
based upon the electroweak gauge group 
$SU(2)_L \times SU(2)_R \times U(1)$ can lead to 
interesting new physics effects 
in the operators $O_{7L}$ and $O_{7R}$.
Due to the extended gauge structure there are both
new neutral and charged gauge bosons, $Z_R$ and $W_R$, as well as
a right-handed gauge coupling, $g_{_R}$.
After the symmetry breaking, the charged $W_R$ mixes with $W_L$ of the SM to
form the mass eigenstates $W_{1,2}$ with eigenvalues $M_{1,2}$. 
And this mixing is described by two parameters; a real mixing angle $\zeta$ 
and a phase $\alpha$,
\begin{eqnarray}
\left( \begin{array}{c} W_1^+ \\ W_2^+ \\ \end{array} \right) =
\left( \begin{array}{cc} \cos ~\zeta & e^{-i \alpha} \sin ~\zeta \\
                        -\sin ~\zeta & e^{-i \alpha} \cos ~\zeta \\ 
       \end{array} \right)
\left( \begin{array}{c} W_L^+ \\ W_R^+ \\ 
       \end{array} \right) .
\end{eqnarray}
In this model the charged current interactions of the right-handed quarks 
are governed by a right-handed CKM matrix $V_R$, which, in principle, 
need not  be related to its left-handed counterpart $V_L$.

If we neglect the charged physical scalar contributions,
the magnetic moment operator coefficients in the LRSM
are given by
\begin{eqnarray}
C_{7L} (m_b) &=& C_{7L}^{SM} (m_b) + A^{tb}
\left[~\eta^{16/23} \tilde{F}(x_t) 
+ {8 \over 3} (\eta^{14/23} -\eta^{16/23}) \tilde{G} (x_t) \right]
+ A^{cb} ~\sum_i~ h_i^\prime \eta^{p_i^\prime} ,
\\
C_{7R} (m_b) &=& (A^{ts})^*
\left[~\eta^{16/23} \tilde{F}(x_t) 
+ {8 \over 3} (\eta^{14/23} -\eta^{16/23}) \tilde{G} (x_t) \right]
+ (A^{cs})^* ~\sum_i~ h_i^\prime \eta^{p_i^\prime} ,
\end{eqnarray}
where
\begin{eqnarray}
C_{7L}^{SM} (m_b) = \eta^{16/23} F(x_t) 
+ {8 \over 3} (\eta^{14/23} - \eta^{16/23}) G(x_t) +
~\sum_i~ h_i \eta^{p_i}, \\
A^{tq} = \zeta~ e^{i \alpha}~ {m_t \over m_b}~ {V_R^{tq} \over V_L^{tq}}~,~~~~~~
A^{cq} = \zeta~ e^{i \alpha}~ {m_c \over m_b}~ {V_R^{cq} \over V_L^{cq}}~,
\end{eqnarray}
with $\eta = \alpha_s(M_{W_1})/\alpha_s(m_b)$ and $x_t = (m_t/m_b)^2$.
The various functions of $x_t$ and 
the coefficients $h_i^{(\prime)}$ and powers $p_i^{(\prime)}$ 
can be found in  Ref. \cite{B2}.
In this paper, we will not constrain ourselves to the LRSM, but discuss the 
general effects of new physics.

Working for the exclusive decay $B \to K^* \ell^+\ell^-$, we need form
factors for the $B\to K^*$ transition.
These form factors can be written \cite{ff} as
\begin{eqnarray}
\langle K^*(p') | \bar s \gamma _\mu b| B (p) \rangle &=& 
ig\varepsilon_{\mu\nu\lambda\sigma}
 \epsilon^{*\nu} (p+p')^\lambda (p-p')^\sigma,\\
\langle K^*(p') | \bar s \gamma _\mu \gamma_5 b|B (p) \rangle &=& 
f  \epsilon^{*}_\mu + a_+ (\epsilon^*\cdot p) (p+p')_\mu 
+a_- (\epsilon^*\cdot p) (p-p')_\mu  ,\\
\langle K^*(p') | \bar s \sigma _{\mu\nu} b| B (p) \rangle &=& 
g_+\varepsilon_{\mu\nu\lambda\sigma} \epsilon^{*\lambda} (p+p')^\sigma+
g_-\varepsilon_{\mu\nu\lambda\sigma} \epsilon^{*\lambda} (p-p')^\sigma\nonumber\\
&&+h\varepsilon_{\mu\nu\lambda\sigma} (p+p')^{\lambda} (p-p')^\sigma 
(\epsilon^*\cdot p),\nonumber\\
\langle K^*(p') | \bar s \sigma _{\mu\nu} \gamma_5 b| B (p) \rangle &=& 
-ig_+\left(\epsilon_{\nu}^*(p+p')_\mu- \epsilon^{*} _\mu (p+p')_\nu\right)
-ig_-\left(\epsilon_{\nu}^*(p-p')_\mu- \epsilon^{*}_\mu (p-p')_\nu\right)
\nonumber\\
&&-2ih\left(p_{\mu}p'_\nu -p_\nu p'_\mu\right) 
(\epsilon^*\cdot p),
\end{eqnarray}
where we have used $\sigma^{\mu\nu}= -\frac{i}{2} 
\varepsilon^{\mu\nu\lambda\sigma} \sigma_{\lambda\sigma} \gamma_5$.
We also use the following definitions,
$\gamma_5= i \gamma^0 \gamma^1 \gamma^2 \gamma^3$ and
$\varepsilon_{0123}=1$. 
The $K^*$ meson subsequently decays to $K$ and $\pi$, with effective
Hamiltonian
\begin{equation}
{\cal H}_{eff} = g_{K^*K\pi} ( p_K -p_\pi) \cdot \epsilon_{K^*}.
\end{equation}
In the following analysis, we neglect the masses of leptons, kaon
and pion. 
The final 4-body
decay amplitude can be written as the sum of two amplitudes,
$$
{\cal A} = {A_R}+{A_L}~, 
$$
where
\begin{eqnarray}
{A_R}=&&
\frac{G_F}{\sqrt{2}} V_{tb}V_{ts}^* g_{K^*K\pi}
 \frac{\alpha m_b}{ \pi L^2} (\bar \ell_R \gamma^\mu l_R)
\left( a_R g_{\mu\nu} -b_R {P_\mu L_{\nu}} +
ic_R ~\epsilon _{\mu \nu \alpha \beta} P^\alpha L^\beta \right ) \\
\nonumber
&&\frac{g^{\nu\alpha}-P^\nu P^\alpha/m_{K^*}^2}
{ P^2-m_{K^*}^2 +i m_{K^*} \Gamma_{K^*}} (p_K-p_\pi)_\alpha,\nonumber \\
{A_L}=&&
\frac{G_F}{\sqrt{2}} V_{tb}V_{ts}^* g_{K^*K\pi}
 \frac{\alpha m_b}{\pi L^2} (\bar \ell_L \gamma^\mu l_L)
\left( a_L g_{\mu\nu} -b_L {P_\mu L_{\nu}} +
ic_L ~\epsilon _{\mu \nu \alpha \beta} P^\alpha L^\beta \right ) \\
\nonumber
&&\frac{g^{\nu\alpha}-P^\nu P^\alpha/m_{K^*}^2}
{ P^2-m_{K^*}^2 +i m_{K^*} \Gamma_{K^*}} (p_K-p_\pi)_\alpha,
\end{eqnarray}
with $P=p_K+p_\pi$, $L=p_+ + p_-$.
The  $a_R,b_R,c_R$ and $a_L,b_L,c_L$ can be expressed as
\begin{eqnarray}
a_L &=& -C_{7-}  \left[2 (P\cdot L) g_+ + L^2 (g_+ +g_-) \right]
-\frac{(C_9 -C_{10}) f}{2m_b}L^2 ,
\label{aaa}\\
b_L &=& -2C_{7-} ( g_+  -  L^2 h )+\frac{(C_9-C_{10}) a_+}{m_b}L^2,\\
c_L &=& -2C_{7+}  g_+ +\frac{(C_9-C_{10}) g}{m_b}L^2,\label{ccc}
\end{eqnarray}
\begin{eqnarray}
a_R &=& -C_{7-}  \left[2 (P\cdot L) g_+ + L^2 (g_+ +g_-) \right]
-\frac{(C_9 + C_{10}) f}{2m_b}L^2 ,
\label{aaar}\\
b_R &=& -2C_{7-} (  g_+ - L^2 h )+\frac{(C_9+C_{10}) a_+}{m_b}L^2,\\
c_R &=& -2C_{7+}  g_+ +\frac{(C_9+C_{10}) g}{m_b}L^2,\label{cccr}
\end{eqnarray}
where $C_{7-} = C_{7R} -C_{7L}$ and $C_{7+} = C_{7R} +C_{7L}$.

The decay rate is computed and the result is
\begin{eqnarray}
\frac{d^5 \Gamma }{dp^2dl^2 d\cos \theta_ K d\cos \theta_+ d \phi}=
\frac{2\sqrt{\lambda}}{128 \times 256 \pi^6 m_B^3}(|A_R|^2+|A_L|^2),
\end{eqnarray}
with $p=\sqrt{P^2}$, $l=\sqrt{L^2}$, and 
$\lambda=(m_B^2-p^2-l^2)^2/4-p^2l^2$. We introduce the various angles  as:
$\theta_K$ is the polar   angle of the
$K$ momentum in the rest system of the $K^*$
meson with respect to the helicity axis, 
{\it i.e.} the outgoing direction of $K^*$.
Similarly $\theta_+$ is the polar angle of 
the positron in the $\gamma^*$
rest system with respect to the helicity axis of the $\gamma^*$.
And  $\phi$ is
the azimuthal angle between the planes of the two decays 
$K^* \to K\pi$ and $ \gamma^* \to \ell^+ \ell^-$.
And then, 
\begin{eqnarray}
|A_R|^2&=&\left|\frac{G_F}{\sqrt{2}} V_{tb}V_{ts}^* g_{K^*K\pi}
 \frac{\alpha m_b}{\pi L^2}\right|^2
 \frac{1}{ (P^2-m_{K^*}^2)^2 + (m_{K^*} \Gamma_{K^*})^2}\nonumber \\ \nonumber
&&\left[|a_R|^2 \left\{ \QL^2-\QN^2-\frac{(L^2-N^2) Q^2}{2} \right\}
\right. \\ \nonumber
&+& 2 Re(a_R b_R^*) \left\{-\QL^2 \PL + \QN \PN \QL \right\} \\ \nonumber
&+& |b_R|^2 \left\{ \PL^2 \QL^2 - \PN^2 \QL^2-\frac{L^2-N^2}{2} P^2 \QL^2 
\right\} \\
\nonumber
&+& |c_R|^2 \left\{- \NPLQ  \NPLQ -\frac{L^2-N^2}{2} \PLQ \cdot \PLQ 
\right\}\\ \nonumber
&+& 2 Im(b_R c_R^*) \PN \QL \NPLQ \\ \nonumber
&+& 2 Im(c_R a_R^*) \QN \NPLQ \\ \nonumber
&-& 2 Im(b_R a_R^*) \QL \NPLQ \\ \nonumber
&-& 2 Re(c_R a_R^*) \LQN \cdot \QPL \\
&+& \left.2 Re(b_R c_R^*) \QL \LPN \cdot \QPL \right],\label{ar}
\end{eqnarray}
and
\begin{eqnarray}
|A_L|^2&=&\left|\frac{G_F}{\sqrt{2}} V_{tb}V_{ts}^* g_{K^*K\pi}
 \frac{\alpha m_b}{\pi L^2}\right|^2
 \frac{1}{ (P^2-m_{K^*}^2)^2 + (m_{K^*} \Gamma_{K^*})^2} \nonumber \\ \nonumber
&&\left[|a_L|^2 \left\{ \QL^2-\QN^2-\frac{(L^2-N^2) Q^2}{2} \right\} 
\right.\\ \nonumber
&+& 2 Re(a_L b_L^*) \left\{-\QL^2 \PL + \QN \PN \QL \right\} \\ \nonumber
&+& |b_L|^2 \left\{ \PL^2 \QL^2 - \PN^2 \QL^2-\frac{L^2-N^2}{2} P^2 \QL^2 
\right\} \\
\nonumber
&+& |c_L|^2 \left\{- \NPLQ  \NPLQ -\frac{L^2-N^2}{2} \PLQ \cdot \PLQ 
\right\}\\ \nonumber
&+& 2 Im(b_L c_L^*) \PN \QL \NPLQ \\ \nonumber
&+& 2 Im(c_L a_L^*) \QN \NPLQ \\ \nonumber
&+& 2 Im(b_L a_L^*) \QL \NPLQ \\ \nonumber
&+& 2 Re(c_L a_L^*) \LQN \cdot \QPL \\
&-& \left.2 Re(b_L c_L^*) \QL \LPN \cdot \QPL \right] , \label{al}
\end{eqnarray}
where $\ABC_{\mu}=
\varepsilon_{{\mu}{\alpha}{\beta}{\gamma}}A^{\alpha}B^{\beta}C^{\gamma}$,
$\ABCD=
\varepsilon_{{\alpha}{\beta}{\gamma}{\delta}}A^{\alpha}B^{\beta}C^{\gamma}D^{
\delta}$, 
$Q=p_K-p_\pi$, and $N=p_+-p_-$.
We use $Tr(\g^{\alpha}\g^{\beta}\g^{\gamma}\g^{\delta}\g_5)=+4 i
\varepsilon^{{\alpha}{\beta}{\gamma}{\delta}} $.
Comparing $|A_L|^2$ with $|A_R|^2$, 
we see that the signs of the corresponding last three terms
are opposite to each other.
We can simplify the expression by introducing the
helicity amplitudes.
The helicity amplitudes are defined as,
\begin{eqnarray}
H_{(\pm1,0)}^L &=& -{\epsilon_{V}^{(\pm,0)}}^{\nu *} {\epsilon_{\gamma}^{(\pm,0)}}^{\mu *}
 \left( a_L g_{\mu\nu} -b_L {P_\mu L_{\nu}} +
ic_L ~\varepsilon _{\mu \nu \alpha \beta} P^\alpha L^\beta \right )~, \nonumber\\
H_{(\pm1,0)}^R &=& -{\epsilon_{V}^{(\pm,0)}}^{\nu *} {\epsilon_{\gamma}^{(\pm,0)}}^{\mu *}
 \left( a_R g_{\mu\nu} -b_R {P_\mu L_{\nu}} +
ic_R ~\varepsilon _{\mu \nu \alpha \beta} P^\alpha L^\beta \right )~.
\label{hel}
\end{eqnarray}
We define the following polarization vectors:
\begin{equation}
\begin{array}{lclccrcl}
\epsilon_V^+ &=& (&0,&1,&i,&0&)/\sqrt{2}~,\\
\epsilon_V^- &=& (&0,&1,&-i,&0&)/\sqrt{2}~,\\
\epsilon_V^0 &=& (&\frac{\sqrt{\lambda}}{m_B},&0,&0,&
\sqrt{\frac{\lambda}{m_B^2}+p^2}&)/p~,\\
\epsilon_{\gamma}^+ &=& (&0,&1,&-i,&0&)/\sqrt{2}~,\\
\epsilon_{\gamma}^- &=& (&0,&1,&+i,&0&)/\sqrt{2}~,\\
\epsilon_{\gamma}^0 &=& (&\frac{\sqrt{\lambda}}{m_B},&0,&0,&
-\sqrt{\frac{\lambda}{m_B^2}+l^2}&)/l~.
\label{pol}
\end{array}
\end{equation}
Substituting them into Eq. (\ref{hel}), we obtain the 
following helicity amplitudes,
\begin{eqnarray}
H_{+1}^L &=&   
\left (a_L+ c_L\sqrt{\lambda} \right)~, \nonumber \\
H_{-1}^L &=&   
\left(a_L- c_L\sqrt{\lambda} \right)~,  \nonumber \\
H_{0}^L &=&  
 -a_L\frac{ P\cdot L}{ pl}+\frac{ b_L\lambda}{pl}~,  \nonumber \\
H_{+1}^R &=&   
\left (a_R+ c_R\sqrt{\lambda} \right)~, \nonumber \\
H_{-1}^R &=&   
\left(a_R - c_R\sqrt{\lambda} \right)~, \nonumber \\
H_{0}^R &=&  
 -a_R\frac{ P\cdot L}{ pl}+\frac{ b_R\lambda}{pl}~.
\end{eqnarray}
Applying the Eqs. (\ref{aaa}-\ref{ccc}), we have
\begin{eqnarray}
H_{+1}^L &=& 2g_+ 
\left (-C_{7-}  (P\cdot L) -C_{7+} \sqrt{\lambda} \right)
-C_{7-}l^2(g_++g_-)-\frac{(C_9-C_{10})l^2}{2m_b}(f-2g\sqrt{\lambda})~,
\nonumber \\
H_{-1}^L &=& 2g_+ 
\left (-C_{7-}  (P\cdot L) +C_{7+} \sqrt{\lambda} \right)
-C_{7-}l^2(g_++g_-)-\frac{(C_9-C_{10})l^2}{2m_b}(f+2g\sqrt{\lambda})~,
\nonumber \\
H_{0}^L &=& \frac{lC_{7-}}{p} \left[ 
2 p^2  g_+ + (P\cdot L) (g_++g_-)+ 2 \lambda  h \right] 
+\frac{(C_9-C_{10})l}{2m_bp}\left[ f(P\cdot L)+2a_+ {\lambda}\right]~,
\label{HS}
\end{eqnarray}
where  $P\cdot L= \sqrt{\lambda+p^2l^2}=
(m_B^2-p^2-l^2)/2$.
The formulae for $H_{+1}^R$, $H_{-1}^R$, $H_{0}^R$ are the same as above 
except that  $C_{10} \mapsto -C_{10}$.
Using the variables $\theta_K$, $\theta_+$, $\phi$, $p$ and $l$, we find:
\begin{eqnarray}
Q\cdot L &=& \sqrt{\lambda} \cos \theta_K~,\nonumber\\
P\cdot N &=& \sqrt{\lambda} \cos \theta_+~,\nonumber\\
P\cdot L &=& \sqrt{\lambda +p^2 l^2} ,\\
Q\cdot N &=& \sqrt{\lambda +p^2 l^2} \cos\theta_K \cos\theta_+ -pl \sin\theta_K
\sin\theta_+ \cos\phi~, \nonumber \\
\NPLQ &=& -pl\sqrt{\lambda} \sin\theta_K \sin\theta_+ \sin\phi~. \nonumber 
\end{eqnarray}
Using these equations, we can get the results for Eqs. (\ref{ar},\ref{al}),
whose sum makes the decay angular distribution of 
$B\to K^*(\to K\pi) \ell^+ \ell^-$,
\begin{eqnarray*}
\frac{d^5 \Gamma }{dp^2dl^2 d\cos \theta_ K d\cos \theta_+ d \phi} &=& 
\frac{
\alpha^2 G_F^2 g_{K^*K\pi}^2 \sqrt{\lambda}p^2 m_b^2 |V_{tb} V_{ts}^*|^2}
{64 \times 
8 (2\pi)^8 m_B^3 l^2 [(p^2-m_{K^*}^2)^2 + m_{K^*}^2 \Gamma_{K^*}^2]} 
\end{eqnarray*}
\begin{eqnarray}
~~~~~~~~&\times &\left\{
4 \cos ^2 \theta_K  \sin^2 \theta_+  (|H_{0}^R|^2+|H_{0}^L|^2) \right.
\nonumber \\
&&+\sin^2 \theta_K  (1+ \cos^2 \theta_+ ) ( |H_{+1}^L|^2+  |H_{-1}^L|^2
+|H_{+1}^R|^2+  |H_{-1}^R|^2) \nonumber \\
&&-2\sin ^2\theta_K \sin^2 \theta_+ \left[ \cos 2\phi Re (H_{+1}^R H^{R*}_{-1}+
H_{+1}^L H^{L*}_{-1} ) \right.
\nonumber \\
&&- \sin 2\phi  \left.Im (H_{+1}^R H^{R*}_{-1}+H_{+1}^L H^{L*}_{-1} ) \right]\nonumber \\
&&-\sin 2\theta_K \sin 2\theta_+ \left[\cos \phi Re (H_{+1}^R H^{R*}_{0}
+H_{-1}^RH_0^{R*}+
H_{+1}^L H^{L*}_{0}+H_{-1}^LH_0^{L*})\right.
\nonumber\\
 &&-\sin \phi \left. Im (H_{+1}^R H^{R*}_{0}-H_{-1}^RH_0^{R*}
+H_{+1}^L H^{L*}_{0}-H_{-1}^LH_0^{L*} ) \right] \nonumber \\
&& -2 \sin^2\theta_K \cos\theta_+ 
(|H_{+1}^R|^2-|H_{-1}^R|^2- |H_{+1}^L|^2+|H_{-1}^L|^2) \nn \\
&&+2 \sin\theta_+ \sin 2\theta_K 
 \left [\cos \phi Re (H_{+1}^R H^{R*}_{0}
-H_{-1}^RH_0^{R*}-H_{+1}^L H^{L*}_{0}
+H_{-1}^LH_0^{L*}) \right. \nn \\
&&-  \sin \phi  \left.\left.
Im (H_{+1}^R H^{R*}_{0}+ H_{-1}^R H^{R*}_{0}
- H_{+1}^L H^{L*}_{0}-H_{-1}^L H^{L*}_{0})
\right] \right\} .
\eea 

If we integrate out the angles $\theta_K$ and $\theta_+$, we get the $\phi$ 
distribution
\begin{eqnarray}
\frac{d \Gamma }{ d \phi} &=& \int
\frac{ 
\alpha^2 G_F^2 g_{K^*K\pi}^2 \sqrt{\lambda}p^2 m_b^2 |V_{tb} V_{ts}^*|^2}
{9\times 16 (2\pi)^8 m_B^3 l^2 [(p^2-m_{K^*}^2)^2 + m_{K^*}^2 \Gamma_{K^*}^2]} 
\left\{
  |H_{0}^R|^2 + |H_{+1}^R|^2+  |H_{-1}^R|^2 \right.\nonumber \\
&&|H_{0}^L|^2 + |H_{+1}^L|^2+  |H_{-1}^L|^2 
-\cos 2\phi Re (H^R_{+1} H^{R*}_{-1}+H_{+1}^L H^{L*}_{-1} )\nonumber \\
&&\left.+ \sin 2\phi Im (H_{+1}^R H^{R*}_{-1}+H^L_{+1} H^{L*}_{-1} )
 \right\} d p^2 d l^2.\label{eqphi}
\end{eqnarray} 
Even if the new physics gives the same total decay rate for $b\to s\gamma$
compared to the SM, {\it i.e.}, we 
cannot see new physics from the $b\to s \gamma$ decay, 
we can still tell new 
physics effects from the angular distribution of $B\to K\pi \ell^+\ell^-$.
If we integrate out the angles $\theta_+$ and $\phi$, we get the $\theta_K$ 
distribution
\begin{eqnarray}
\frac{d \Gamma }{ d \cos\theta_K} &=& \int
\frac{(2 \pi)
\alpha^2 G_F^2 g_{K^*K\pi}^2 \sqrt{\lambda}p^2 m_b^2 |V_{tb} V_{ts}^*|^2}
{3\times 64 (2\pi)^8 m_B^3 l^2 [(p^2-m_{K^*}^2)^2 + m_{K^*}^2 \Gamma_{K^*}^2]} 
\left\{ 2 \cos^2 \theta_K (
  |H_{0}^R|^2 \right.\nonumber \\
&&+\left. |H_{0}^L|^2)+ \sin^2\theta_K \left(|H_{+1}^R|^2+  |H_{-1}^R|^2+
|H_{+1}^L|^2+  |H_{-1}^L|^2 \right)
 \right\} d p^2 d l^2.\label{eqtheta}
\end{eqnarray}
Taking the narrow resonance limit of $K^*$ meson, {\it i.e.},
using the  equations
\begin{eqnarray}
&&\Gamma_{K^*}=\frac{g_{K^*K\pi}^2 m_{K^*}}{48 \pi}~, \nn \\
&&\lim_{\Gamma_{K^*} \to 0} 
\frac{\Gamma_{K^*} m_{K^*}}{(p^2-m_{K^*}^2)^2 + m_{K^*}^2 \Gamma_{K^*}^2}=
\pi \delta(p^2-m_{K^*}^2)~,
\end{eqnarray}
we can perform the integration over $p^2$ and obtain the
double differential branching ratio with respect to 
dilepton mass squared $l^2$ and azimuthal angle $\phi$, 
\begin{eqnarray}
{dBr \over dl^2 d \phi}&=& \tau_{B}{\alpha^2 {G_F}^2 \over 384 \pi^5}
\sqrt{\lambda} {{m_b}^2 \over {m_B}^3 l^2} |V_{ts}V_{tb}|^2 {1 \over 2 \pi}
\left\{
  |H_{0}^R|^2 + |H_{+1}^R|^2+  |H_{-1}^R|^2 \right.\nonumber \\
&&+|H_{0}^L|^2 + |H_{+1}^L|^2+  |H_{-1}^L|^2 
-\cos 2\phi Re (H^R_{+1} H^{R*}_{-1}+H_{+1}^L H^{L*}_{-1} )\nonumber \\
&&\left.+ \sin 2\phi Im (H_{+1}^R H^{R*}_{-1}+H^L_{+1} H^{L*}_{-1} )
 \right\}, 
\end{eqnarray}
and
\begin{eqnarray}
 {dBr \over d{l^2}}&=&\tau_{B}{\alpha^2 {G_F}^2 \over 384 \pi^5}
\sqrt{\lambda} {{m_b}^2 \over {m_B}^3 l^2} |V_{ts}V_{tb}|^2
\left\{
  |H_{0}^R|^2 + |H_{+1}^R|^2+  |H_{-1}^R|^2 \right.\nonumber \\
 &&\left. +|H_{0}^L|^2 + |H_{+1}^L|^2+  |H_{-1}^L|^2 \right\}, \label{38}
\end{eqnarray}
where $\tau_{B}$ is the life time of $B$ meson, and 
we replace all $p$ by $m_{K^*}$ due to the
$\delta$ function. 

We further define the distribution $r(\phi, \hat{s})$ as
\begin{eqnarray}
&&r(\phi, \hat{s}) \nn \\ 
&\equiv& \left[{dBr \over d{l^2} d \phi}\right]/
\left[{dBr \over d{l^2} }\right] \nn \\
&=&{1 \over 2 \pi} \left\{
1-\frac{\cos 2\phi Re (H^R_{+1} H^{R*}_{-1}+H_{+1}^L H^{L*}_{-1} )
- \sin 2\phi Im (H_{+1}^R H^{R*}_{-1}+H^L_{+1} H^{L*}_{-1} )}
{|H_{0}^R|^2 + |H_{+1}^R|^2+  |H_{-1}^R|^2 
 +|H_{0}^L|^2 + |H_{+1}^L|^2+  |H_{-1}^L|^2} \right\}, 
\label{R}
\end{eqnarray}
where $\hat{s}=l^2 /m_B^2$. The distribution
$r(\phi,\hat{s})$ is the probability for finding $K$ meson per unit 
radian region in the direction of azimuthal angle $\phi$.
Therefore $r(\phi,\hat{s})$ oscillates around its average value given by
${1 \over 2 \pi}\simeq 0.16$. 


\section{Numerical Analyses and Conclusions}

In the numerical calculations, we use
the form factors calculated in Ref. \cite{ff2}. 
They are listed in Table~ \ref{t1} for zero momentum transfer. 
The revolution formula for these form factors is 
\begin{equation}
f_i (l^2) = \frac{f_i (0)}{ 1- \sigma_1 l^2 +\sigma_2 l^4}~,
\end{equation} 
where $l^2 = (p_{\ell^+} + p_{\ell^-})^2$.
The corresponding values $\sigma_1$ and $\sigma_2$ for each form factors 
are also listed in Table 1.

The analytic Wilson coefficients $C_7^{eff}(\mu)$, $C_9^{eff}(\mu)$, and
$C_{10}(\mu)$ in the SM are 
given in Ref. \cite{buras}.
Under the leading logarithmic approximation, we get the numerical results 
\cite{KMS} at $\mu=m_b$:
\begin{equation}
C_7^{eff}=-0.311, ~~~C_{10}=-4.546,
\end{equation}
and to the next-to-leading order,
\begin{eqnarray}
C_9^{eff}&=& 4.153
+0.381~ g(m_c/m_b,\hat s) +0.033~ g(1,\hat s) +0.032~g(0,\hat s) ,\label{c9}
\end{eqnarray}
where $\hat s=l^2/m_b^2$.
The function  $g(z,\hat s)$ can be found in Ref. \cite{buras}.
Here for numerical evaluation, 
we use $m_{top}=175$ GeV, $m_b=4.8$ GeV,
$m_c=1.4$ GeV, $\Lambda_{QCD}=214$ MeV.
We include the $J/\psi$ contribution as done in \cite{long},
\begin{equation}
C_9^{eff}\to C_9^{\prime eff}=C_9^{eff}-C^{0}~ \kappa 
\frac{ 3 \pi \Gamma[\psi \to
\ell^+ \ell^-] 
m_{\psi}}{ \alpha^2  (l^2 -m_\psi^2 
+im_\psi\Gamma_\psi)}\; .
\end{equation}
where  $\kappa=2.3$ and $C^{(0)}=0.381.$

The decay width for inclusive $b \to  s \gamma$ decay 
in terms of operators $O_{7L}$ and $O_{7R}$ is given by
\begin{equation}
\Gamma (b \to  s \gamma) = { G_F^2 m_b^5 \over 32 \pi^4 } \alpha_{em}
|V_{ts}^* V_{tb}|^2 \left( |C_{7L} (m_b)|^2 + |C_{7R} (m_b)|^2 \right) .
\end{equation}
It is convenient to normalize this radiative partial width to
the semileptonic rate
\begin{equation}
\Gamma (b \to  c e \bar{\nu}) =
{ G_F^2 m_b^5 \over 192 \pi^3 } |V_{cb}|^2 f({m_c / m_b}) 
\left[ 1 - {2 \over 3 \pi} \alpha_s (m_b) g({m_c / m_b}) \right] ,
\end{equation}
where $f(x)=1-8 x^2-24 x^4 \ln x +8 x^6-x^8$ represents a phase space factor, 
and the function $g(x)$ encodes next-to-leading order QCD correction
effects \cite{cskim}.
In terms of the ratio $R$, 
\begin{equation}
R \equiv {\Gamma (b \to  s \gamma) 
\over \Gamma(b \to  c e \bar{\nu})} =
{6 \over \pi} {|V_{ts}^* V_{tb}|^2 \over |V_{cb}|^2}
{\alpha_{em} \over f({m_c / m_b})}
{ |C_{7L} (m_b)|^2 + |C_{7R} (m_b)|^2 \over 
1 - {2 \over 3 \pi} \alpha_s (m_b) g({m_c / m_b}) } ,\label{rr}
\end{equation}
the $b \to  s \gamma$ branching fraction is obtained by
\begin{equation}
{\cal BR}(b \to  s \gamma) 
\simeq {\cal BR}(B \to X_c l \nu)_{\rm exp.} \times R
\simeq (0.105) \times R .
\end{equation}
For ${\cal BR}(b \to  s \gamma)$, 
we use the present experimental value \cite{CLEO} of the branching 
fraction for $B \to  X_s \gamma$ decay, 
\begin{equation}
{\cal BR}(B \to  X_s \gamma) = 
(3.15 \pm  0.35 \pm 0.32 \pm 0.26) \times 10^{-4} .
\end{equation}
Constrained by this experiment, we derive from Eq. (\ref{rr})
\begin{equation}
|C_{7L} (m_b)|^2 + |C_{7R} (m_b)|^2=0.081\pm 0.014.\label{c7}
\label{constraint}
\end{equation}
In general, we can parameterize $C_{7L}$ and $C_{7R}$ as follows
by introducing parameters $(x,~u,~v)$,
\begin{eqnarray}
C_{7L}&=&-\sqrt{|C_{7L} (m_b)|^2 + |C_{7R} (m_b)|^2} \cos{x} \exp{i(u+v)}~,\nn \\
C_{7R}&=&\sqrt{|C_{7L} (m_b)|^2 + |C_{7R} (m_b)|^2} \sin{x} \exp{i(u-v)}~,
\label{para}
\end{eqnarray}
where $u$ is a common phase of $C_{7L}$ and $C_{7R}$, and $v$ denotes 
the relative phase between $C_{7L}$ and $C_{7R}$.
In Figs. 1--6, 
we show the distribution $r(\phi, \hat{s})$ 
for different sets of $(x, u, v)$.
The minimum of the invariant mass is set to be $0.7$ GeV in the
figures.
We can understand the qualitative features 
in the region  of small invariant mass
by comparing with an approximate formula for the azimuthal 
angle distribution.  
By using Eq. (\ref{para}), 
we  can show that in the small invariant mass
limit, $r(\phi,\hat{s})$ defined in Eq. (\ref{R})
is written as,
\begin{eqnarray}
r(\phi,\hat{s}) &\simeq&{1 \over 2 \pi} \left\{
1+\cos 2\phi \frac{Re (C_{7R} C_{7L}^*)}{|C_{7R}|^2+|C_{7L}|^2}
- \sin 2\phi \frac{Im (C_{7R} C_{7L}^*)}{|C_{7R}|^2+|C_{7L}|^2}
\right\} \nn \\
&=&{1 \over 2 \pi}\left\{1-\frac{1}{2}
\sin 2 x \cos 2(\phi-v)\right\}.
\label{approx}
\end{eqnarray}
The equation follows from the fact that   
the helicity amplitudes are dominated by 
the two coefficients $C_{7R}$ and $C_{7L}$ in the region of low invariant 
mass,  
\begin{eqnarray}
H^{L,R}_{+1}& \simeq&-4 g_+ C_{7R} \sqrt{\lambda}, \nn \\
H^{L,R}_{-1}& \simeq& 4 g_+ C_{7L} \sqrt{\lambda},\nn \\
H^{L,R}_{0}& \simeq &0.
\end{eqnarray}

The SM case ($C_{7R} \simeq 0$) 
corresponds to $(x, u, v)=(0,0,0)$, and 
$r(\phi,\hat{s})$ is shown in Fig. 1. In the SM there are only small 
phase shifts from the $g(z,\hat s)$ in Eq. (\ref{c9}) \cite{buras},
which are practically negligible because of 
$Im (H_{+1}^R H^{R*}_{-1}+H^L_{+1} H^{L*}_{-1} )\simeq 0$. 
The last term of (\ref{R}) vanishes for any $\hat s$. It is shown
in Fig. 1 that there is only  $(-\cos 2 \phi)$ behavior for larger $\hat s$.
We can also note that as $\hat{s}$ is getting smaller,
the $\phi$ dependence even vanishes. This is consistent with 
formula (\ref{approx}), since $x=0$ in the SM. 
Another extreme case, $C_{7L}=0$, is shown in 
Fig. 2. There still remains $\phi$ dependence even in low invariant mass
region. We checked that $\phi$ dependence vanishes by going further to
smaller invariant mass $l \ll 1$ GeV, which is not shown in the 
figure. This shows that there is large contribution from $C_9$ and $C_{10}$
even for rather low invariant mass $l \sim 1$ GeV. 
For  larger $\hat s$, near 0.4, there is some disorder appearing in Fig. 2.
It represents the interference effect of the short distance contribution
with the long distance contribution from $J/\psi$ resonance.

If  $\frac{|C_{7R}|}{|C_{7L}|} =O(1)$, the approximate formula (\ref{approx})
works qualitatively well [see Figs. 3--6].   
There we change the relative phase 
of $C_{7R}$ and $C_{7L}$ by setting  $\frac{|C_{7R}|}{|C_{7L}|} =1$.
In Fig. 3, $u=v=0$, then there is no imaginary part. 
We can read from Fig. 3 the 
$(-\cos 2 \phi)$ behavior for $\frac{C_{7R}}{C_{7L}}=-1$ 
in the region of  small $\hat s$.
For  larger $\hat s$, there is interference from the $C_9$ and $C_{10}$
contributions, and the resulting figure is not so simple.
From Fig. 4, we can see the $(+\cos 2 \phi)$
behavior for  $\frac{C_{7R}}{C_{7L}}=1$ in the  region of small $\hat s$. 
This is consistent with the approximate
formula (\ref{approx}). It is the 
inverse case of Fig. 3. Finally we introduce CP
violating phase $v$ between
$C_{7L}$ and $C_{7R}$, which leads to the phase shifts. In Figs. 5 and 6,
we choose $v=\pm \pi / 8 $. According to Eq. (\ref{approx}),
it amounts to $\pm \pi / 4$ in the phase shift, which
can be seen in Figs. 5 and 6.
We do not show figures for non-zero values of $u$, which is the relative 
phase between $C_{7i}$'s and $C_9$ ($C_{10}$). The non-zero value of this
angle $u$ will not change the $r(\hat s, \phi)$ behavior at low $\hat s$ 
[see Eq. (\ref{approx})], but will change it at higher $\hat s$.
This area is affected by the interference of $C_{7i}$'s and 
$C_9$ ($C_{10}$).

Using eq.(\ref{38}), we do the integration with $l$ from 0.4 GeV to 1.2 GeV,
we get the branching ratio of $B\to K\pi \ell^+\ell^-$ at this region:
$ 1 \times 10^{-7}$. From the figures we know that in the above region,
it is effective to distinguish the new physics contribution. 
The number of B mesons we need is around $10^{10}$, which can not be 
produced in the current B factories, but possible in the future LHC-B etc.
More concretely, dividing the region of $\phi$ into 10 bins,
we expect $10^2$ events in each bin in the standard model.
If the distribution follows from the formulae Eq.(\ref{approx}) 
with $\sin 2x =1$,
the numbers of the event of each bin are no more flat and it
oscillates between 50 and 150. If this is the case,
we can surely distinguish the distribution from the flat one
of the standard model.

To summarize, we studied the angular distribution of 
$B\to K^*(\to K \pi) \ell^+\ell^- $. 
We showed that the azimuthal angle $(\phi)$ distribution is very useful
for probing possible new physics effects and for confirming
the SM through this flavor-changing neutral current process. 
Here  $\phi$ is the angle between the decay plane of $(K \pi)$ and
the decay plane of $(\ell^+ \ell^-)$.
In particular, if the two operators $O_{7L}$ and $O_{7R}$,
which contribute to $B \to K^* \gamma$, 
are equally important, then the $\phi$ dependence is significant.
In the SM case, there is only a weak $(-\cos 2 \phi)$ dependence 
for the region of small $\hat s$, but the term proportional to 
$(-\cos 2 \phi)$ becomes dominant  for the region of  larger $\hat s$.
When new physics is introduced without changing the decay rate of the
$b \to s \gamma$, we can nontheless have quite different angular distribution for
$B \to K\pi\ell^+\ell^-$.
We also showed that the phase shift results in the appearance of $(\sin 2\phi)$
term, the latter-thus being a clear signature of the presence of CP violating phase. 
Even if we cannot probe new physics from $B \to K^* \gamma$,
it is possible to see the new physics effects through the azimuthal angle 
distribution of $B \to K\pi\ell^+\ell^-$.
We also note that $C_9$ and $C_{10}$ are about 
ten times larger than $C_7$'s. Therefore, even in the region of small dilepton mass,
their effect cannot be neglected. In our analysis, their
effect has been fully incorporated. 

\section*{Acknowledgement}

We thank G. Cvetic for careful reading of the manuscript and his
valuable comments.
The work of C.S.K. was supported 
in part by  grant No. 1999-2-111-002-5 from the Interdisciplinary 
Research Program of the KOSEF,
in part by the BSRI Program, Ministry of Education, Project No.
99-015-DI0032, 
and in part by Sughak program of 
Korean Reasearch Foundation, Project No. 1997-011-D00015.
The work of Y.G.K. is supported by KOSEF Postdoctoral Program.
The work of T.M. is supported by Grant-in-Aid for Scientific Research
on Priority Areas (Physics of CP violation) and the work of C.L.
is supported by JSPS.


\begin{table}[hb]
\caption{Form factors in zero momentum transfer and parameters of
revolution formula [11].}
\begin{center}
\begin{tabular}{c|ccccccc}
\hline
            & $g$  &  $f$ &  $a_+$  & $a_-$  &  $g_+$  & $g_-$  &  $h$ \\
$f_i(0)$    &0.063 & 2.01 & -0.0454 & 0.053  & -0.3540 & 0.313  & -0.0028 \\
\hline
$\sigma_1$ & 0.0523  & 0.0212  & 0.039 & 0.044 & 0.0523 & 0.053 & 0.0657\\
$\sigma_2$ & 0.00066 & 0.00009 &0.00004&0.00023& 0.0007 &0.00067& 0.0010\\
\hline
\end{tabular}\end{center}\label{t1}
\end{table}

\begin{figure}\begin{center}
\epsfig{file=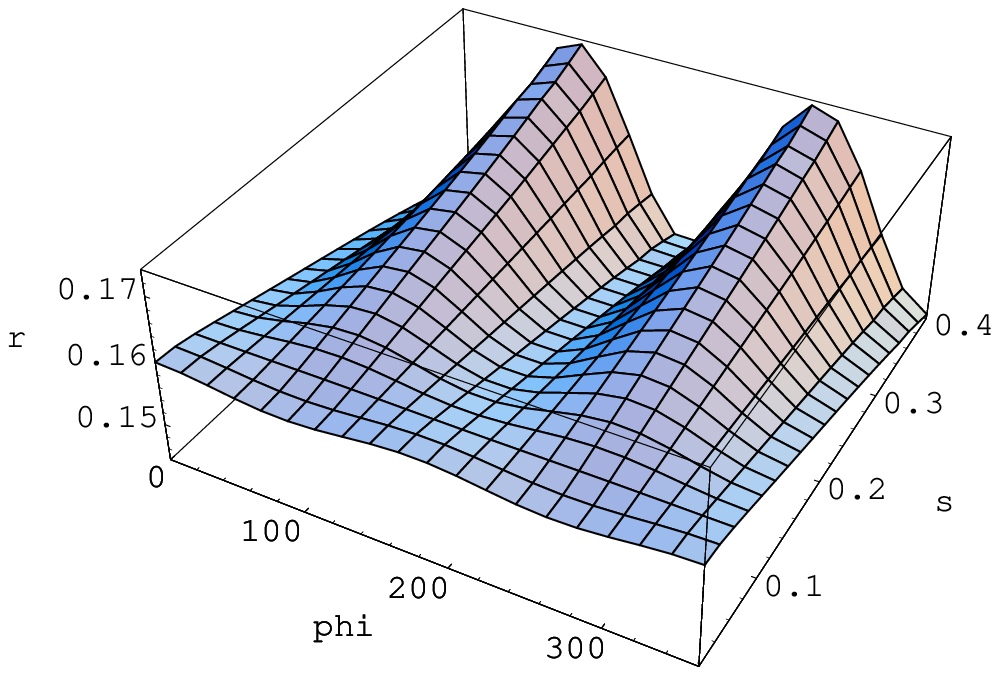, 
width=11.5cm,angle=0}
\caption{The distribution $r(\phi, \hat{s})$ for $(x,u,v)=(0,0,0)$,
{\it i.e.}, $\frac{C_{7R}}{C_{7L}}=0$.
This case corresponds to the standard model case.
Here $\phi$ is the azimuthal angle between the decay plane of $(K\pi)$
and the decay plane of $(\ell^+ \ell^-)$, and 
$\hat{s} = (p_{\ell^+} + p_{\ell^-})^2/m_b^2$.} 
\end{center}
\end{figure}

\begin{figure}\begin{center}
\epsfig{file=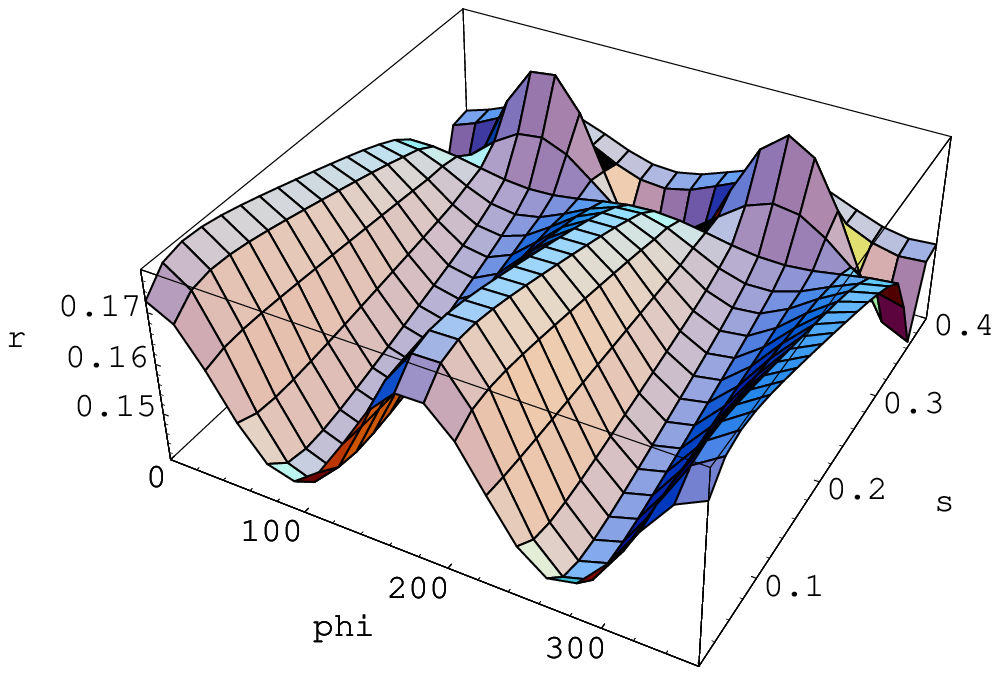, 
width=11.5cm,angle=0}
\caption{The distribution $r(\phi, \hat{s})$ for $(x,u,v)=(\pi/2,0,0)$,
{\it i.e.}, $\frac{C_{7L}}{C_{7R}}=0$.}
\end{center}
\end{figure}

\begin{figure}\begin{center}
\epsfig{file=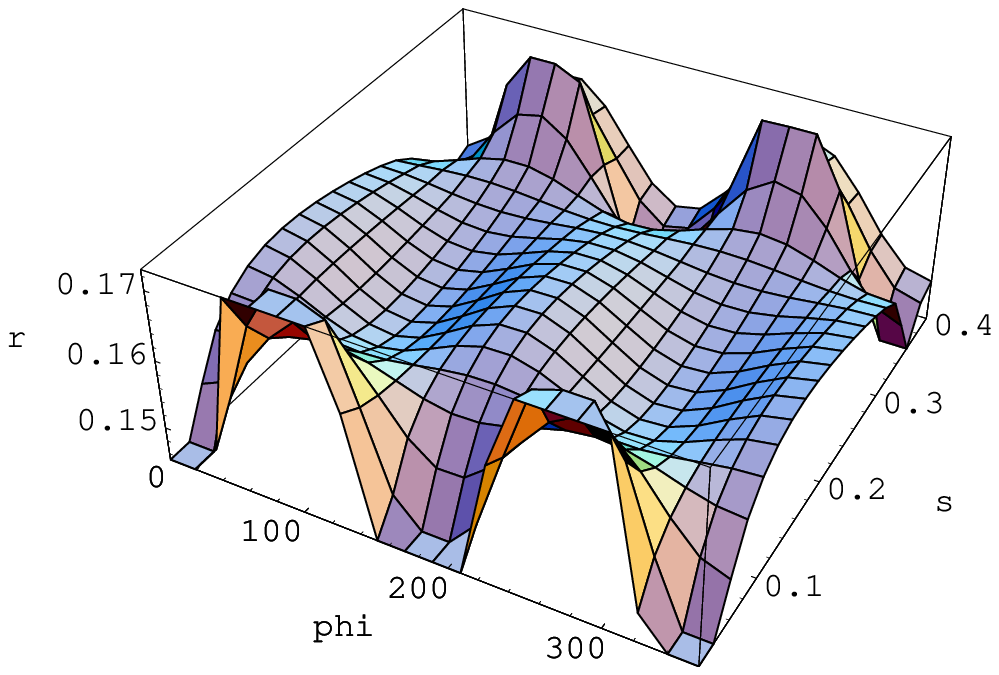, 
width=11.5cm,angle=0}
\caption{The distribution $r(\phi, \hat{s})$ for $(x,u,v)=(\pi/4,0,0)$,
{\it i.e.}, $\frac{C_{7R}}{C_{7L}}=-1$.}
\end{center}
\end{figure}

\begin{figure}\begin{center}
\epsfig{file=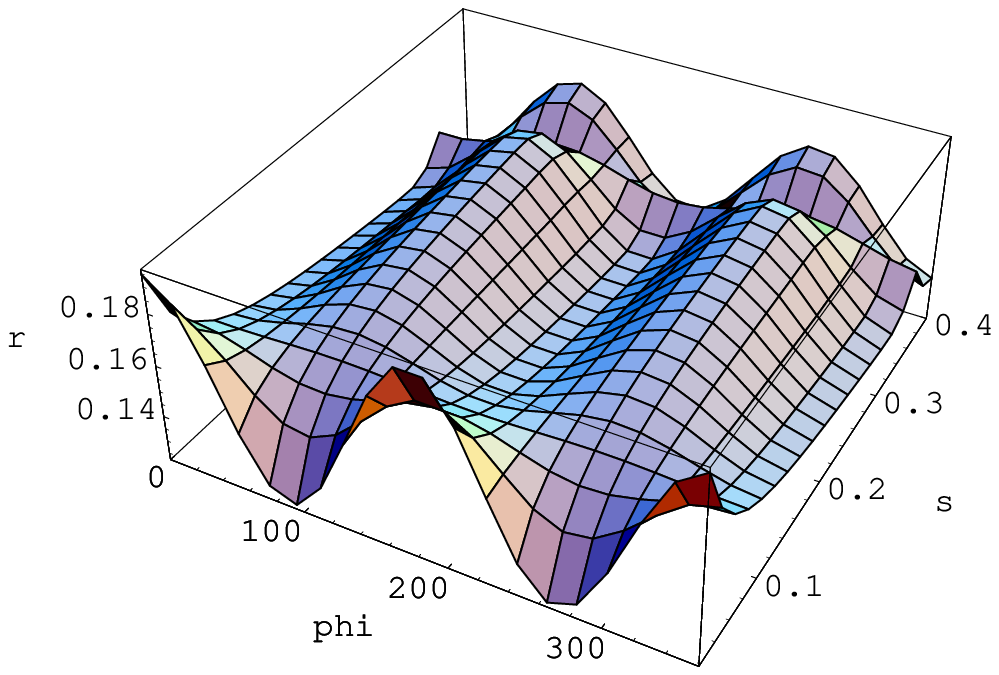,
width=11.5cm,angle=0}
\caption{The distribution $r(\phi, \hat{s})$ for $(x,u,v)=(-\pi/4,0,0)$,
{\it i.e.}, $\frac{C_{7R}}{C_{7L}}=+1$.}
\end{center}
\end{figure}

\begin{figure}\begin{center}
\epsfig{file=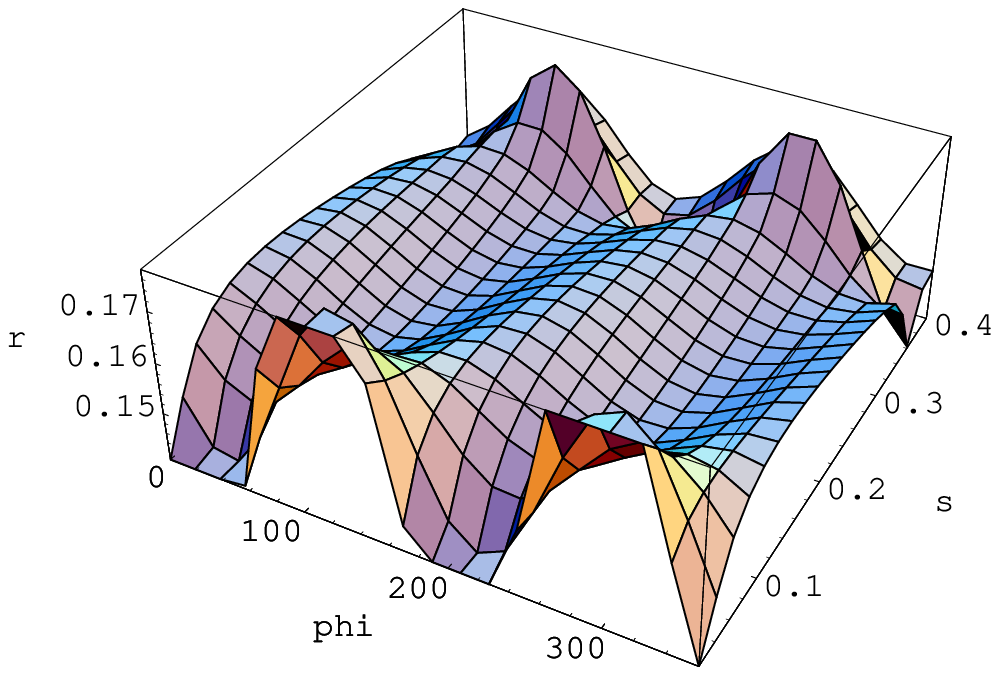,
width=11.5cm,angle=0}
\caption{The distribution $r(\phi, \hat{s})$ for $(x,u,v)=(\pi/4,0,\pi/8)$,
{\it i.e.}, $\frac{C_{7R}}{C_{7L}}=-\exp(-i\pi/4)$.}
\end{center}
\end{figure}

\begin{figure}\begin{center}
\epsfig{file=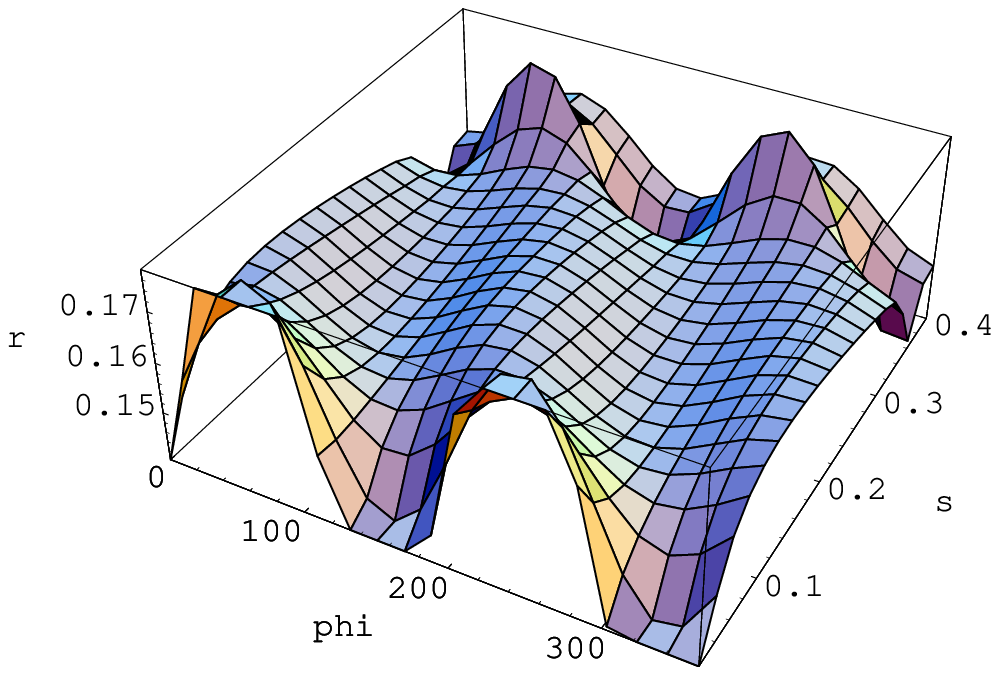,
width=11.5cm,angle=0}
\caption{The distribution $r(\phi, \hat{s})$ for $(x,u,v)=(\pi/4,0,-\pi/8)$,
{\it i.e.}, $\frac{C_{7R}}{C_{7L}}=-\exp(i\pi/4)$.}
\end{center}
\end{figure}
\end{document}